\documentclass[pra,aps,preprint]{revtex4}

\usepackage{amssymb}
\usepackage{epsfig}
\usepackage{graphicx}
\begin{document}
\title{Repulsive Casimir forces and the role of surface modes}
\author{I. Pirozhenko$^{1}$}
\author{A. Lambrecht$^{2}$}
\affiliation{$^1$Bogoliubov Laboratory of Theoretical Physics, JINR, 141980
Dubna, Russia\\
$^2$Laboratoire Kastler Brossel, CNRS, ENS,
UPMC - Campus Jussieu case 74, 75252 Paris, France}
\pacs{42.50.Ct,12.20.Ds,12.20.-m}
\begin{abstract}
The Casimir repulsion between a metal and a dielectric suspended in
a liquid has been thoroughly studied in recent experiments. In the present paper
we consider  surface modes in three layered systems modeled by dielectric functions
guaranteeing repulsion. It is shown  that surface modes play a decisive role in this
phenomenon at short separations. For a toy plasma model we find the contribution
of the surface modes at all distances.
\end{abstract}
\date{\today}
\maketitle

\section{Introduction}
The existence of repulsive Casimir forces follows straightforwardly
from the Lifshitz theory~\cite{Lifshitz:Eng:1956} for a special
choice of the materials. The phenomenon has been discussed in
several theoretical
papers~\cite{Boyer:1974,Kenneth:2002ij,Henkel:2005,Pirozhenko:2008tr,Dalvit:2008}
mainly for materials with nonunit magnetic permeability. Recent
advances in the design of metamaterials~\cite{Veselago:2006}
demonstrating nontrivial magnetic permeability at optical
frequencies stimulated experiments seeking after the Casimir
repulsion. Though some experimental groups have reported noticeable
decrease of the attraction~\cite{Binns:2008} in the presence of
metamaterials, to our knowledge, nobody has succeeded so far in
reversing the sign of the force.

The repulsive Casimir forces  between {\textit purely dielectric} materials
now appear to be more promising for observation and might
turn out useful for nanomechanical systems.  There repulsion is achieved by filling
the space between the bodies with a medium with a precisely chosen
dielectric permittivity.

It was experimentally shown~\cite{Milling:1996, Lee:2001,Lee:2002} for short separations ($<$50 nm) that
repulsive van der Waals forces between specially chosen dielectrics with intervening liquids of low-polarity
(eg. cyclohexane , ethanol, or bromobenzene) agree with theoretical predictions of the Lifshitz theory including the retardation contribution.  The repulsive Casimir forces  at large separations up to several hundred nanometers  have
not been experimentally studied until very recently~\cite{Rodriguez:2008,Munday:2007,Munday:2008,Capasso:2007,Capasso:2009}

The setup is the following. Two surfaces '1' and '3' with
dielectric permittivities $\varepsilon_1$ and $\varepsilon_3$ are
separated by a gap filled by medium '2' with dielectric permittivity
$\varepsilon_2$. In recent experiments or proposals medium '2' is a
liquid, ethanol~\cite{Capasso:2007,Palasantzas:2009a} or
bromobenzene~\cite{Capasso:2009} while the two surfaces are made of
silica and gold. In order to obtain a repulsive force the
respective dielectric functions have to satisfy the relation
\begin{equation}
\varepsilon_1(i\omega)<\varepsilon_2(i\omega)<\varepsilon_3(i\omega).
\label{condition}
\end{equation}
in the frequency range relevant for the force measurement.

The Casimir force is given
by~\cite{Lifshitz:Eng:1956,Lambrecht:2000}
\begin{equation}
F(L)=-\frac{\hbar}{2\pi^2}\sum_{\rho}\int\limits_0^{\infty} dk
k \int\limits_0^{\infty} d\omega \kappa_2\,\frac{r^{\rho}_{12}\, r^{\rho}_{32}}{\exp(2\kappa_2
L)-r^{\rho}_{12}\, r^{\rho}_{23}}.
\label{eq1}
\end{equation}
Here $L$ is the width of the gap between the plates;
$r^{\rho}_{i2}(i\omega,\kappa)$, $\rho=TE,TM$, are the reflection
coefficients at imaginary frequencies for the surfaces facing the
medium '2'
\begin{equation}
r^{TM}_{i2}(i\omega)=\frac{\varepsilon_2\kappa_i-\varepsilon_i \kappa_2}{\varepsilon_2\kappa_i+\varepsilon_i
\kappa_2},\quad r^{TE}_{i2}(i\omega)=-\frac{\kappa_i-\kappa_2}{\kappa_i+
\kappa_2},
\label{eq2}
\end{equation}
with  $\kappa_j(i\omega)=\sqrt{k^2+\varepsilon_j(i\omega)\omega^2/c^2}$,
 $j=1,2,3$.

In the present system the Casimir force is repulsive.
Indeed, at short distances, $L<<\lambda_{p,i}$, the force is approximated by  $F\approx-H_{123}/3L^3$. Its
magnitude and sign is defined by the non-retarded Hamaker constant~\cite{Bergstrom:1997}
\begin{eqnarray}
H_{123}&=&\frac{3\hbar}{8\pi^2}
\int\limits_0^{\infty} d\omega\sum_{n=1}^{\infty}\frac{(\triangle_{12}[i\omega]\,\triangle_{32}[i\omega])^n}{n^3}.
\label{A13}
\end{eqnarray}
with $\triangle_{j2}=(\varepsilon_2(i\omega)-\varepsilon_j(i\omega))/(\varepsilon_2(i\omega)+\varepsilon_j(i\omega))$.
Under the condition (\ref{condition}) the Hamaker constants are negative~\cite{Milling:1996,Capasso:2007} and the force is repulsive.

At long distances $L>>\lambda_{p,i}$, where the largest contribution to the force
comes from small frequencies and small wave vectors $\kappa$ the reflection coefficients tend to their the static values
\begin{equation}
r^{TE}_{j2},r^{TM}_{j2}\approx\lim_{\omega\to0}
\frac{\sqrt{\varepsilon_2(i\omega)}-\sqrt{\varepsilon_j(i\omega)}}{\sqrt{\varepsilon_2(i\omega)}+
\sqrt{\varepsilon_j(i\omega)}}.
\quad j=1,3.
\end{equation}
Plugging them into (\ref{eq1}) one gets a rough estimation of the
force at large plates' separations~\cite{pirozhenko:013811}. It is
easy to check numerically that if the materials obey
(\ref{condition}), especially at low frequencies, the force is
repulsive in the long distance limit. These considerations help to
choose the materials for the devices based on repulsive Casimir
forces~\cite{Rodriguez:2008,Capasso:2009}. However the origin of
this repulsion itself is not quite clear.

In the following we will consider this question in relation with
surface modes. The electromagnetic quantum fluctuations which give
rise to the Casimir effect obey the Maxwell equations. The
corresponding boundary-value problem has two types of solutions: the
propagative waves, which will be called photonic modes in the
following, and the waves living on the interfaces and exponentially
decaying outwards. These surface modes exist only in the TM
polarization. Further on we call them surface plasmons or polaritons
depending on the model which describes the material. The term
"plasmon" is reserved for the plasma model. The expression
(\ref{eq1}) comprises the contributions from both propagative and
surface modes. The present paper pursues their respective
roles in the Casimir repulsion by considering a model system
consisting of three layers of dielectric material the plasma
frequencies of which are chosen such that condition \ref{condition}
is fulfilled. Section II and III describe the definition of the
surface modes and their contribution to the Casimir energy. In
section IV we perform a calculation of the repulsive Casimir force
for the recent experiment \cite{Capasso:2009} where we use as
dielectric functions the two oscillator model for the silica surface
and the bromobenzene liquid filling the gap. The second surface is
covered with gold for which we use the Drude model as dielectric
response. A comparison between this calculation and the predictions
of our model system allows to define the limitations of the latter.
The paper finishes with some conclusive remarks in section V.

We will assume that the dielectric properties of the liquid do not
change as its layer becomes thicker or thinner. We will not consider
the hydrodynamics of this system either. Of cause, a body moving in
a liquid due to the Casimir attraction or repulsion experiences drug
force, which depends on the velocity. This hydrodynamic force should
be taken into account in the experiments. Luckily it does not depend
on the reason of the movement, and may be estimated
separately~\cite{Munday:2007, Munday:2008, Capasso:2009}.

\section{The interaction of the surface plasmons}
Let us first formulate the interaction of the surface modes to
understand the nature of Casimir repulsion. For simplicity we
consider materials described by plasma model,
$\varepsilon_i=1-\omega^2_{p i}/\omega^2$, $i=1..3$, where
$\omega_{p i}$ is the material's plasma frequency. The frequency of
the single surface plasmon living on the interface of medium $i$,
$i=1,3$, with medium '2' is given by
\begin{equation}
\omega^{sp}_{i2}=\frac{1}{\sqrt{2}}\left[2 k^2 c^2+\omega_{p2}^2+\omega_{pi}^2-\sqrt{4 k^4 c^4+(\omega_{p2}^2-
\omega_{pi}^2)^2}\right]^{\frac{1}{2}}
\label{plasm_nonid}
\end{equation}
When $k\to 0$ the single surface plasmon frequency tends to
$\omega_{p2}$, provided $\omega_{pi}/\omega_{p2}<1$, otherwise it
approaches $\omega_{pi}$. In the limit $k\to\infty$ it tends to
$\sqrt{\omega_{pi}^2+\omega_{p2}^2}/\sqrt{2}$, $i=1,3$.

The surface modes evanescent in the direction of the gap between the plates are coupled through
the cavity.  Their frequencies are defined by the equation
\begin{equation}
\prod\limits_{i=1,3}^{}\frac{\varepsilon_2 q_i+\varepsilon_i q_2}{\varepsilon_2 q_i-\varepsilon_i q_2}=e^{-2\,q_2\,L},
\label{eq_plasm}
\end{equation}
with $q_i=\sqrt{k^2-\varepsilon_i \omega^2/c^2}$. Let us compare the
behavior of the solutions of Eq.~(\ref{eq_plasm}) in two cases which
result in opposite signs of the Casimir force.
Fig.1
\begin{figure}[tb] \epsfig{file=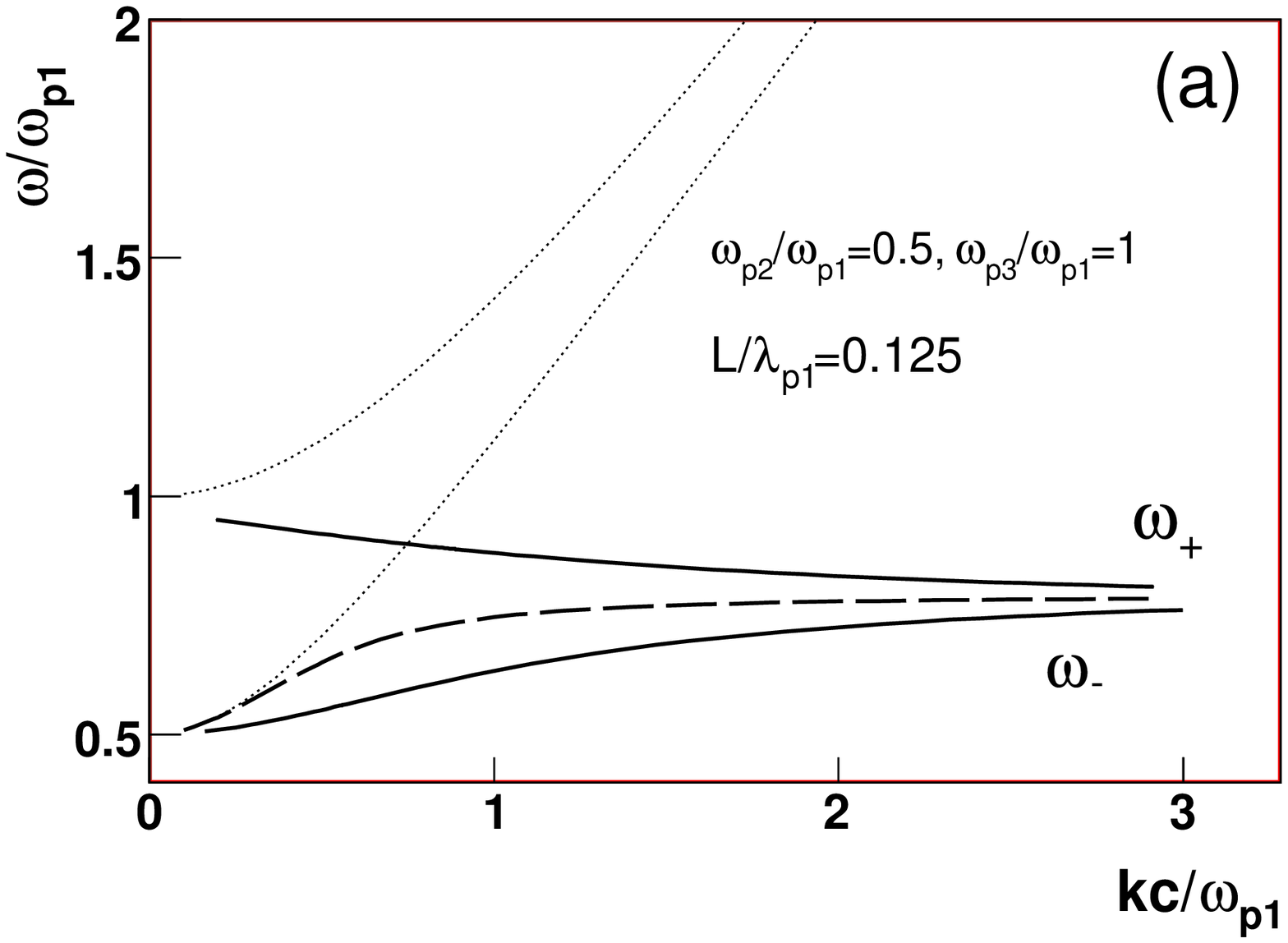,width=8.0cm}
\epsfig{file=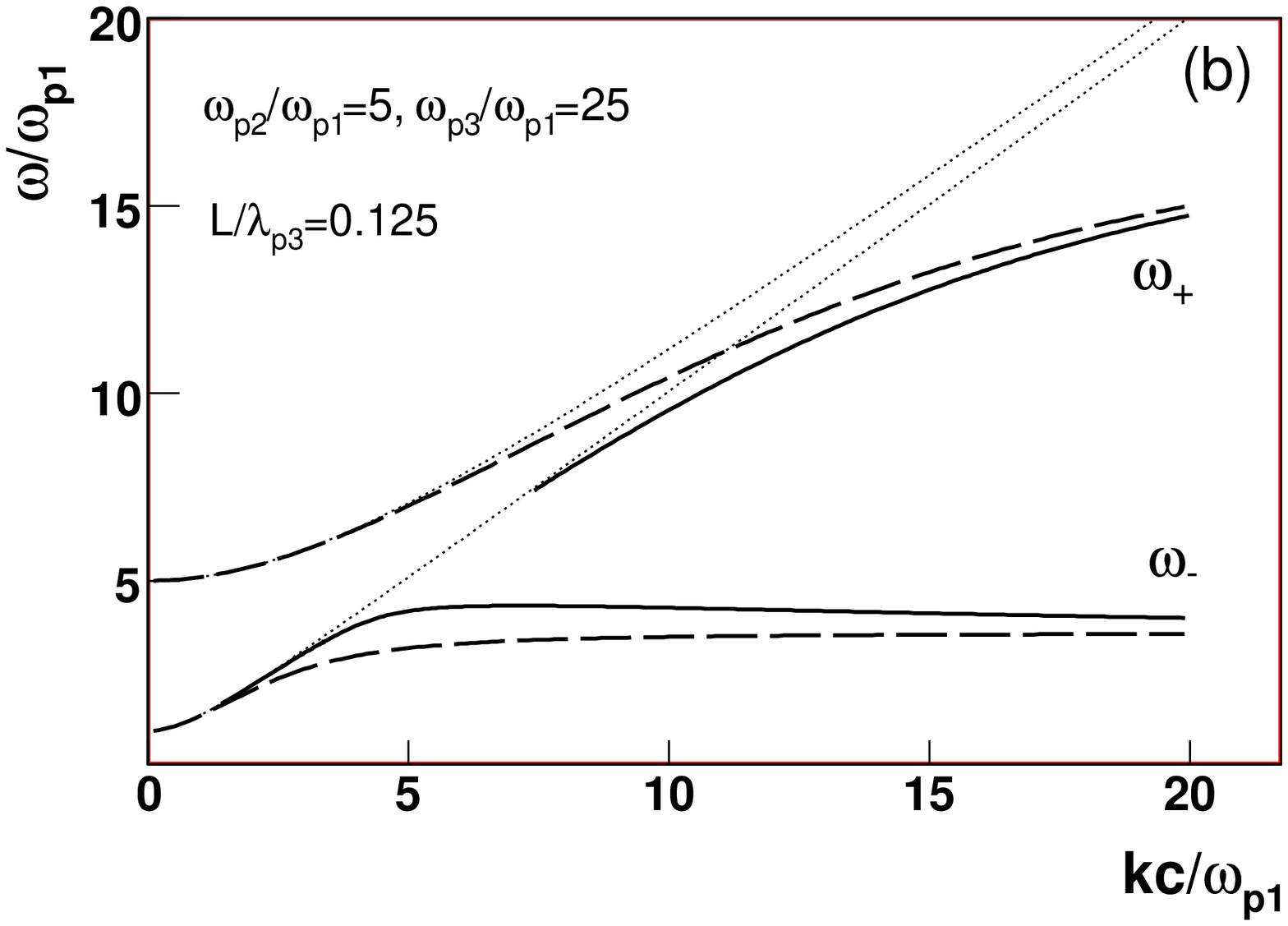,width=8.0cm} \caption{The plasmon
modes in two set-ups: (a) $\omega_{p2}<\omega_{p1},\omega_{p3}$
(attractive Casimir force), (b)
$\omega_{p1}<\omega_{p2}<\omega_{p3}$ (repulsive Casimir force). The
solid lines correspond to coupled symmetric and antisymmetric
plasmons. The dashed lines show the single surface plasmons living
on the interfaces 1-2 and 2-3. The dotted lines mark the boundaries
between propagative and evanescent sectors.} \label{fig1}
\end{figure}
shows the plasmon modes inside the "sandwich" (a)
$\omega_{p2}<\omega_{p1},\omega_{p3}$ (attractive Casimir force),
(b) $\omega_{p1}<\omega_{p2}<\omega_{p3}$ (repulsive Casimir force).
The coupled plasmon modes are plotted as solid lines. The dashed
lines correspond to the plasmons living on the single interfaces.
Situation (a) corresponds to well known case where the gap in-between
the two plates is filled with vacuum fluctations ($\omega_{p2}=0$).
It has been studied in detail in the papers~\cite{IntraLam:2005,
Bordag:2005}. For the sake of comparison with case (b)
let us briefly recall the results. Fig.1a shows the solutions of
equation~(\ref{eq_plasm}) when
$\alpha\equiv\omega_{p2}/\omega_{p1}=0.5,\;
\beta\equiv\omega_{p3}/\omega_{p1}=1$ (equal slabs interlaid with
material having $\omega_{p,2}$). The dotted lines starting from
$\omega/\omega_{p1}=\alpha$ and $\beta$ separate the propagative and
evanescent sectors respectively in the gap and in the slabs. We
denote  the solutions of (\ref{eq_plasm}) by $\omega_{\pm}$, where
$\omega_{+}$ is usually referred to as antisymmetric plasmon and
$\omega_-$ as symmetric plasmon. The latter one lies entirely in the
evanescent sector , $k^2-\omega^2/c^2+\omega_{p2}^2/c^2>0$. In
contrast, the antisymmetric plasmon $\omega_{+}$ penetrates into the
propagative sector if $k<p_{+}$,
\begin{equation}
p_{+}=\frac{\omega_{p1}}{c}\left\{\frac{\beta^2+\frac{\sqrt{\beta^2-\alpha^2}}{\sqrt{1-\alpha^2}}+
\alpha^2\sqrt{\beta^2-\alpha^2}\;\Lambda}{1+\frac{\sqrt{\beta^2-\alpha^2}}{\sqrt{1-\alpha^2}}+
\sqrt{\beta^2-\alpha^2}\;\Lambda}-\alpha^2\right\}^{\frac{1}{2}},
\label{bound1}
\end{equation}
where $\Lambda=L\omega_{p1}/c$ is the dimensionless distance. The
coupled plasmons $\omega_{\pm}$ surround the single surface plasmon
solution $\omega^{sp}$, and $\omega_{+}|_{k\to0}=\omega_{p1}$,
$\omega_{-}|_{k\to0}=\omega_{p2}$. At large wave vectors
$\omega_{-}\to\omega^{sp}_{12},\; \omega_{+}\to\omega^{sp}_{32}$.
For equal slabs, meaning $\beta=1$, these limits coincide (dashed
line). The modes which are evanescent in the gap are evanescent
within the slabs as well, because on Fig.1a the borderline between
the evanescent and the propagative sectors in the gap lies below the
corresponding borderline for the slabs.

Fig.1b shows just the opposite situation. The dotted lines starting
from $\omega/\omega_{p1}=1$ and $\alpha$ separate the propagative
and evanescent sectors respectively in the gap and in the slab with
$\omega_{p1}$. Now the borderline between the two sectors referring
to the gap lies above the one corresponding to the slab. In the area
between these curves the equation~(\ref{eq_plasm}) has no real
solutions as $q_2^2, q_3^2>0$, but $q_1^2<0$.
Solving~(\ref{eq_plasm}) with
$k^2-\omega^2/c^2+\omega_{p2}^2/c^2\to0$, one finds  that the mode
$\omega_{+}$ exists for $k>k_{+}$,
\begin{eqnarray}
k_{+}&=&\frac{\omega_{p1}}{c}\left[\frac{\alpha^2+\beta^2\,f}
{1+f}-1\right]^\frac{1}{2}, \label{bound2}\\
f&=&\frac{\sqrt{\alpha^2-1}}{\sqrt{\beta^2-1}}\,\tanh{(\Lambda\sqrt{\alpha^2-1})}.
\nonumber
\end{eqnarray}
The coupled plasmons $\omega_{\pm}$  lie inside the area enveloped by the single surface plasmon
solutions $\omega^{sp}_{32}$ and $\omega^{sp}_{12}$ defined by~(\ref{plasm_nonid}), and
$\omega_{-}|_{k\to0}=\omega_{p1}$, $\omega_{+}|_{k\to0}=\omega_{p2}$.

\section{The vacuum energy of the surface modes}
In the following we will concentrate on the Casimir energy
contribution of surface modes in the case of a repulsive force. The
renormalized vacuum energy of the interacting surface plasmons
living on the plane mirrors is formally given by
\begin{equation}
E^{sp}=\frac{\hbar}{2}\sum_{\sigma}^{}\int\limits_{k_{(\sigma)}}^{\infty}\frac{d
k\, k}{2\pi} \left[\omega_{\sigma}\right]^{L}_{L\to\infty}, \quad \sigma=\pm,
\label{energy_plasm}
\end{equation}
where $\lim_{L\to\infty}\omega_{+}=\omega^{sp}_{12}$,  $\lim_{L\to\infty}\omega_{-}=\omega^{sp}_{32}$, $k_{-}=0$, and $k_{+}$ is defined by~(\ref{bound2}).

To calculate the surface plasmon energy, we first introduce the
dimensionless variables $K=kL, \Omega=\omega L/c$, and
$\Omega_{p1}\equiv\Lambda$, $\Omega_{p2}=\alpha\Lambda$,
$\Omega_{p3}=\beta\Lambda$ in~(\ref{eq_plasm},\ref{energy_plasm})
\begin{equation}
E^{sp}=\frac{\hbar c}{4 \pi L^3}\left\{\int\limits_{K_{+}}^{\infty} d
K K (\Omega_{+}-\Omega_{12}^{sp})+\int\limits_{0}^{\infty} dK K(\Omega_{-}-\Omega_{32}^{sp})\right\}.
\label{energy_plasm_dimless}
\end{equation}
Then we change the integration variable in
(\ref{energy_plasm_dimless}), $K\rightarrow Q=\sqrt{K^2-\Omega^2}$
and write the renormalized energies of the symmetric and
antisymmetric plasmons as
\begin{eqnarray}
\bigl.E^{sp}&=&\frac{\hbar c}{4 \pi L^3}\left\{\frac{1}{2}\int\limits_{-\Lambda^2}^{\infty} dQ^2 \{\Omega_{-}+\Omega_{+}-\Omega_{12}^{sp}-\Omega_{23}^{sp}\}\right.\nonumber\\
&-&\left.\frac{1}{2}\int\limits_{X}^{-\Lambda^2}dQ^2  \Omega_{12}^{sp}
+\frac{1}{3}(\Omega_{+}^3-\Omega_{12}^3)\bigl|_{K_{(+)}}^{\infty}\bigr.\right\},\label{sp_long}
\end{eqnarray}
where $X=K_+^2-\Omega_{12}^2(K_+)$.
\begin{figure}[ttb]
\epsfig{file=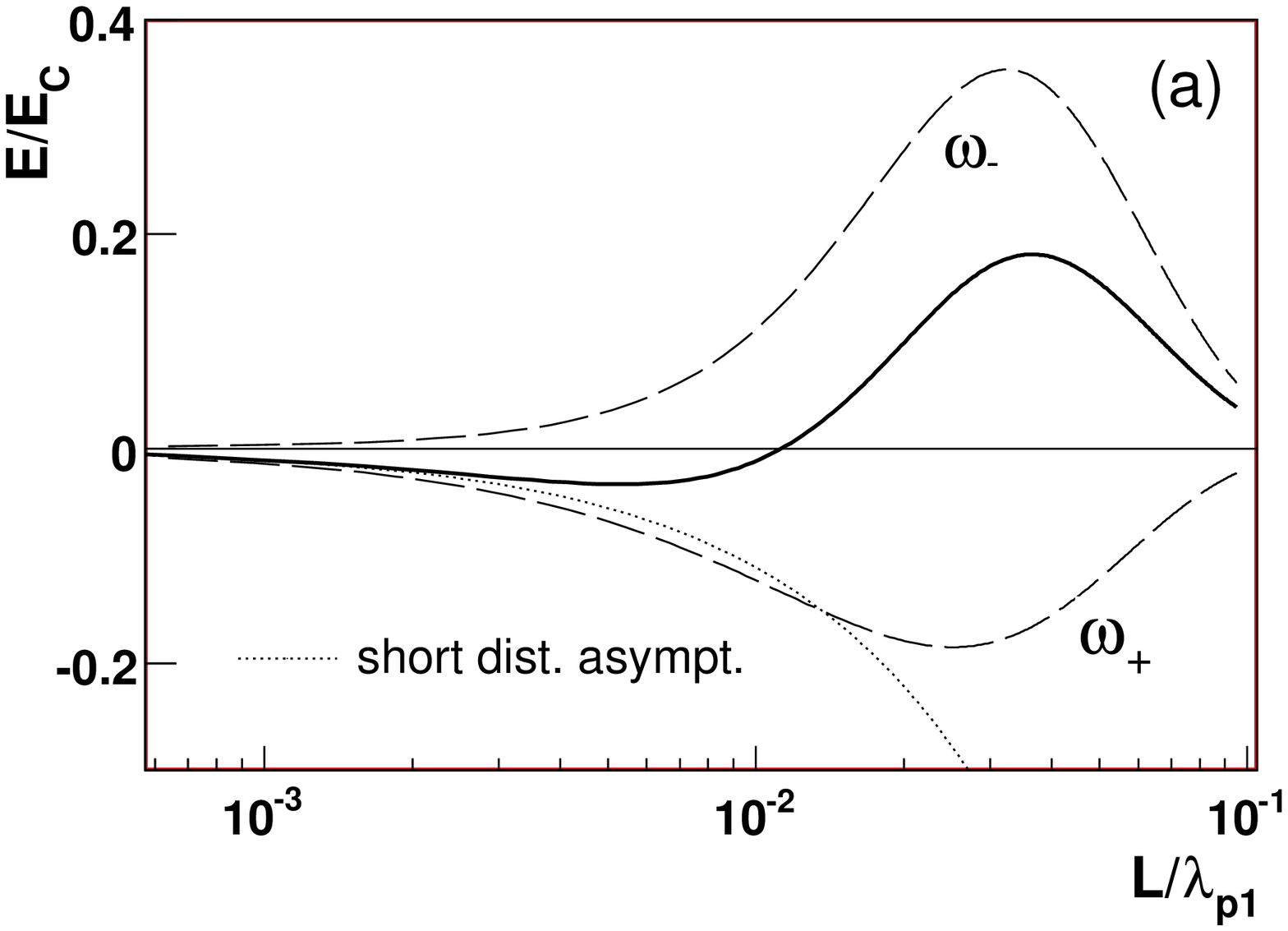,width=8.0cm}
\epsfig{file=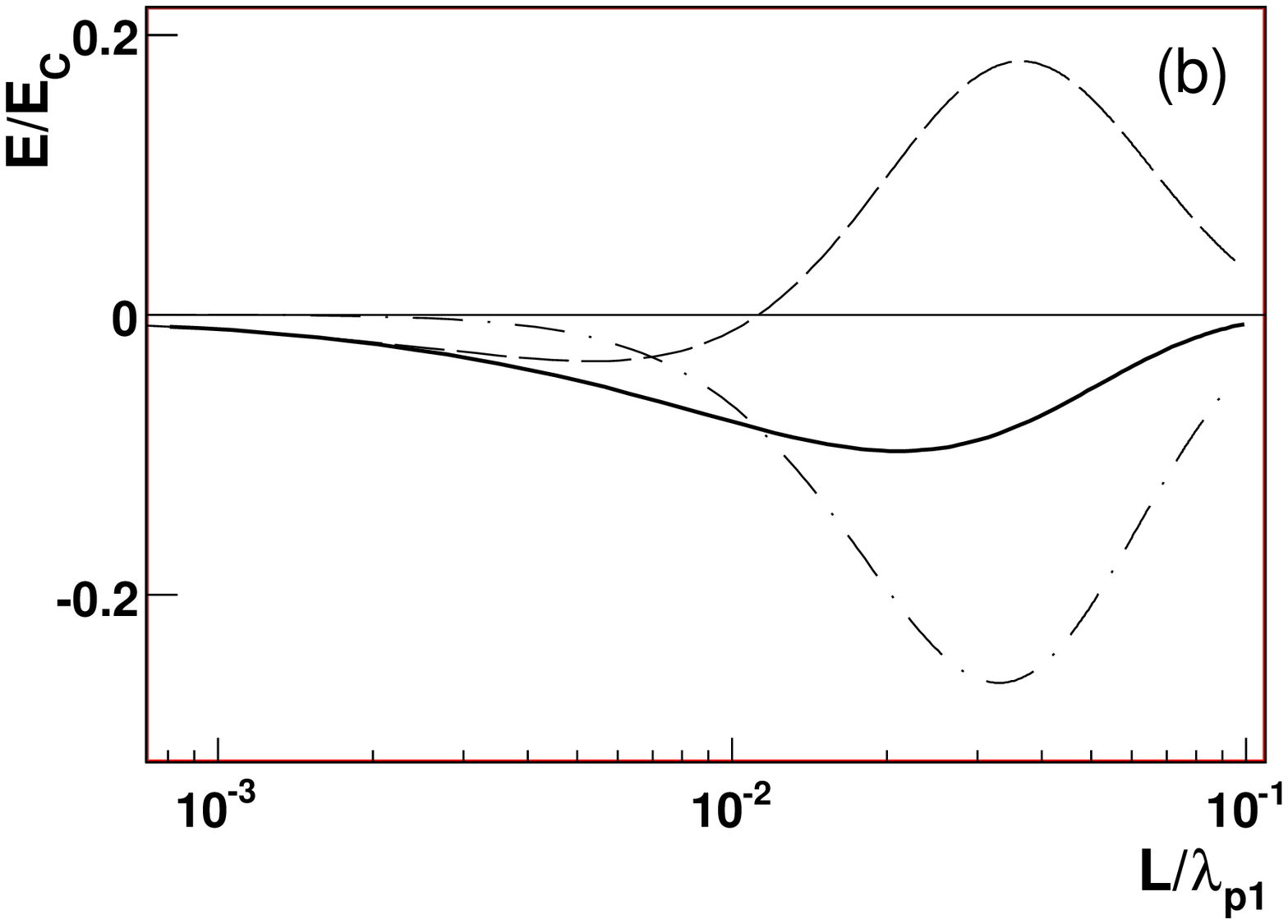,width=8.0cm} \caption{The
normalized vacuum energy: (a) symmetric plasmon $\omega_{-}$ (dashed
line), antisymmetric plasmon $\omega_{+}$ (dash-dotted line), sum of
both (solid line); (b) plasmon energy (dashed line), photon energy
(dash-dotted line), total Casimir energy (solid line).
$\omega_{p2}/\omega_{p1}=5, \omega_{p3}/\omega_{p1}=25$.}
\label{fig2}
\end{figure}

The numerical results for the plasmon energy are given in
Fig.\ref{fig2}(a). The distance is normalized by the largest plasma
wavelength in the system, $\lambda_{p1}=2\pi c/\omega_{p1}$. The
plot shows the reduction factor which is the ratio of the plasmon
vacuum energy and the Casimir energy of perfectly conducting plates,
$E_C=-\hbar c\pi^2/720L^3$. As the Casimir force between perfectly
conducting plates is attractive, a negative or positive reduction
factor corresponds to repulsion or attraction respectively. For the
energies, the arguments goes in the opposite way. The energy of the
antisymmetric plasmon $\omega_+$ is positive, corresponding to
repulsion with a negative reduction factor, while the energy of the
symmetric plasmon $\omega_-$ is negative yielding attraction and
thus a positive reduction factor. The antisymmetric plasmon
dominates at short distances, while the symmetric plasmon makes
decisive contribution at larger separations. Thus the total plasmon
interaction is repulsive at short separations and attractive at
medium and long distances. Here we estimate the length scale with
respect to $\lambda_{p1}$.

In \cite{Bordag:2005, IntraLam:2005}, it was shown that for two
plates separated by vacuum the attractive Casimir energy is a result
of cancelations between plasmon and photon contributions at all
distances. Moreover the plasmon and photon energies have different
signs and decrease more slowly with distance  ($\sim L^{-5/2}$) than
the total energy ($\sim L^{-3}$).

Fig.\ref{fig2}(b) shows the reduction factor for the total Casimir
energy when condition~(\ref{condition}) is met. Clearly it is
negative at all distances corresponding to a repulsive Casimir
interaction. It comes out as the sum of negative photon contribution
and the plasmon contribution, plotted in Fig.\ref{fig2}(a). The
plasmon contribution dominates at short separations. At medium and
long distances  the photon contribution prevails assuring repulsion.
At distances large with respect to $\lambda_{p1}$ the  energy
decreases as $E\sim \textrm{Ei}(1,2\alpha\Lambda)$, $\Lambda=2\pi
L/\lambda_{p1}$. The decrease is much steeper than in any set-up
yielding attraction.

The attraction of the surface plasmons at short ranges explains the
sign and the magnitude of the Casimir force between plates separated
by vacuum (or any material with
$\omega_{p2}<\omega_{p1},\omega_{p3}$). In the following we will
show that under the condition (\ref{condition}) the surface plasmons
produce a repulsive contribution at short distances.

The equation for the interacting surface plasmons may be solved
explicitly for large wave vectors which correspond to short
distances:
\begin{eqnarray}
(\omega_{\pm})^2=\frac{1}{4}\bigl\{2\omega_{p,2}^2+\omega_{p,1}^2+\omega_{p,3}^2 \bigr.
\mp[(\omega_{p,1}^2-\omega_{p,3}^2)^2 \nonumber\\
+4 (\omega_{p,1}^2-\omega_{p,2}^2)(\omega_{p,3}^2-\omega_{p,2}^2)
e^{-2 |k| L}]^{1/2} \}.
\label{plasm_nonid_short}
\end{eqnarray}
Then the vacuum energy is
\begin{equation}
E^{sp}_{as}=\frac{\hbar \omega_{p,1}}{32 \pi L^2}Y(\alpha,\beta)
\label{energy_plasm_short}
\end{equation}
where
\begin{eqnarray}
Y&=&\int\limits_{0}^{\infty}dk k \{
\psi_{+}+\psi_{-}-\sqrt{2}[(1+\alpha^2)^{\frac{1}{2}}+(\beta^2+\alpha^2)^{\frac{1}{2}}]\},\nonumber \\
\psi_{\pm}&=&\left\{2\alpha^2+1+\beta^2\mp[(1-\beta^2)^2\right.\nonumber\\
&&\left.+4 (1-\alpha^2)
(\beta^2-\alpha^2)e^{-k}]^{\frac{1}{2}}\right\}^{\frac{1}{2}}.
\label{energy_plasm_short1}
\end{eqnarray}
If at large frequencies condition (\ref{condition}) is satisfied,
then $E^{sp}_{as}>0$ yielding repulsion. The simplest particular
case is $\beta=\alpha^2>1$
($\omega_{p,2}/\omega_{p,1}=\omega_{p,3}/\omega_{p,2}$)
corresponding to equal ratios between the plasma frequencies. For
large $\alpha$, $\psi(\alpha\to\infty)\to 0.67 \alpha$. If
$\alpha=1+\delta$, $Y(\delta\to0)\approx\delta^2/2$. For example,
$Y(1.1)=0.00497$. Thus, the vacuum energy of the interacting surface
plasmons is positive corresponding to repulsion.

\section{Repulsive Casimir force using the dielectric response of the materials}
\begin{figure}[bt]
\epsfig{file=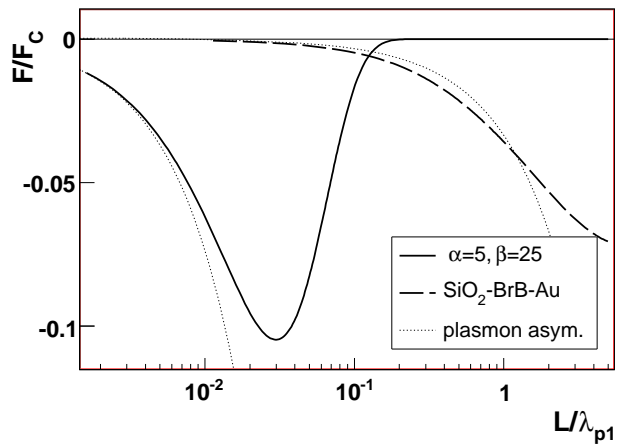,width=9.0cm}
\caption{Reduction factor $\eta_F=F/F_C$ at short plate separations as a function of dimensionless distance $\Lambda= L/\lambda_{p1}$; $\alpha=\omega_{p,3}/\omega_{p,2}=\omega_{p,2}/\omega_{p,1}$.  The dotted lines are the respective short distance asymptotes of the surface plasmon interaction. $\alpha=\omega_{p,3}/\omega_{p,2}=\omega_{p,2}/\omega_{p,1}$. For the system SiO$_2$-C$_6$H$_5$Br-Au  $\lambda_{p1}=2\pi c/\omega_{p,SiO_2}^{eff}=88.44$nm.}
\label{fig3}
\end{figure}

Let us now compare the repulsive force due to the interaction of the surface modes obtained
in the plasma model with the corresponding total Casimir force calculated according to
the zero temperature Lifshitz formula~(\ref{eq1}).  On Fig.\ref{fig3} we plot the
force~(\ref{eq1}) normalized by the Casimir force between perfect
conductors as a function of the dimensionless distance
$L/\lambda_{p1}$, $\lambda_{p1}=2\pi c/\omega_{p,1}$. The
calculation is done for tree layers with
$\varepsilon_i=1+\omega_{pi}^2/\omega^2$,
$\alpha=\omega_{p2}/\omega_{p1}=5$,
$\beta=\omega_{p3}/\omega_{p1}=25$ (solid line). At short distances
the force coincides with the short distance asymptote of the
interacting surface plasmons, $F^{sp}_{as}=-dE^{sp}_{as}/dL$,
$F^{sp}_{as}/F_C=-7.38 L/\lambda_{p1}$ (dotted line). At long
distances ($L>>\lambda_{p2}$) the force is repulsive and decays much
faster than the force between two plates separated by vacuum,
$F\sim\exp(-2\omega_{p2}L/c)/L$,  where $\omega_{p2}$  is the plasma
frequency of the material filling the gap. This behavior is
explained by the divergence of $\varepsilon_2$ for $\omega\to0$
within the plasma model. If dielectric function of the intermediate
material '2' is finite at low frequencies, the reduction factor
saturates at large distances.

An example of a more realistic set-up is the system Silica-Bromobenzene-Gold.
Repulsive dispersion forces in a similar system
were measured by~\cite{Milling:1996} at short distances, the results appeared to be consistent with the calculated non-retarded Hamaker constants. Precisely the same system was studied in a recent experiment~\cite{Capasso:2009}
at separations from 20 nm to several hundred nanometers. Here we do not consider separations
larger than 300 nm, so that we need not to account for temperature corrections.
In the following we first  calculate  the reduction factor applying the zero temperature
Lifshitz formula~(\ref{eq1}) with the materials described by Drude and Lorentz models. 

For gold $\varepsilon_{Au}(i\omega)=1+\omega_{Au}^2/[\omega(\gamma+\omega)]$,
where $\omega_{Au}=1.367\cdot10^{16}$rad/s,
$\gamma=5.316\cdot10^{13}$rad/s~\cite{Lambrecht:2000}. To simplify the analysis for
bromobenzene and silica we confine ourselves to two oscillators in
the standard multiple oscillator model
\begin{equation}
\varepsilon_i(i\omega)=1+\frac{C_{IR}^i}{1+(\omega/\omega_{IR}^i)^2}+\frac{C_{UV}^i}{1+(\omega/\omega_{UV}^i
)^2},
\end{equation}
where $i=1,2$ and correspond to silica and bromobenzene
respectively. The various parameters are $C_{IR}^1=0.829$,
$\omega_{IR}^1=0.867\cdot10^{14}$rad/s, $C_{UV}^1=1.098$,
$\omega_{UV}^1=2.034\cdot10^{16}$ rad/s~\cite{Bergstrom:1997};
$C_{IR}^2=2.967$, $\omega_{IR}^2=5.47\cdot10^{14}$rad/s,
$C_{UV}^2=1.335$, $\omega_{UV}^2=1.286\cdot10^{16}$
rad/s~\cite{Milling:1996}. With these values of the parameters the
condition~(\ref{condition}) is satisfied for
$\omega<9\cdot10^{15}$rad/s, where
$\varepsilon_{SiO_2}<\varepsilon_{C_6H_5Br}<\varepsilon_{Au}$. For
$\omega>3\cdot10^{16}$rad/s,
$\varepsilon_{Au}<\varepsilon_{C_6H_5Br}<\varepsilon_{SiO2}$. Both
regions contribute to repulsion {\it within the two oscillator model
for bromobenzene and silica}. The applicability of this model is
discussed in~\cite{Capasso:2009,Palasantzas:2009a}.

In the present situation the solutions of~(\ref{eq_plasm}) are
interacting surface polaritons. At close separations realized by a
 thin layer of bromobenzene, where the decisive contribution to the
force comes from large wave vectors and high frequencies, the
absorbtion is negligible, and the models used for the materials are
reduced to the plasma model with effective plasma frequency
$\omega_{p
i}^{eff}=[C_{UV}^i\omega_{UV}^2+C_{IR}^i\omega_{IR}^2]^{1/2}_i$,
$i=1,2$, $\omega_{p3}=\omega_{Au}$. We find  $\omega_{
SiO_2}^{eff}=2.131\cdot10^{16}$rad/c,  $\omega_{
C_6H_5Br}^{eff}=1.488\cdot10^{16}$rad/c.

At close separation the surface polaritons turn into surface
plasmons, and the
formulae~(\ref{plasm_nonid_short}-\ref{energy_plasm_short1}) for the
the interacting surface plasmons become valid. But one should keep
in mind that they describe correctly only short distance regime.
Substituting $\alpha=\omega_{p2}^{eff}/\omega_{p1}^{eff}=0.698$,
$\beta=\omega_{p3}/\omega_{p1}^{eff}=0.641$, we get
$F^{sp}_{as}/F_C=-0.03355\, L/\lambda_{p1}^{eff}$, where
$\lambda_{p1}^{eff}\equiv\lambda_{SiO_2}^{eff}=2\pi
c/\omega_{p1}^{eff}=88.44$nm is the effective plasma wavelength for
the composed system. In this regime the reduction factor for the
interacting surface modes is negative yielding repulsion. It
coincides with the total reduction factor for the force  at short
distances.

At medium and low frequencies the dielectric functions of the
materials are complex, and the calculation of the vacuum energy
corresponding to the surface  modes is not so straightforward. The
total reduction factor for the force is shown as the dashed line in
Fig.3. Clearly repulsion sets in for distances of the order of or
larger than the effective plasma frequency while the toy model gives
repulsion essentially for distances much smaller than that and gives
a negligible force for larger distances. In contrast at large
distances our calculation using the two oscillator model yields a
negative force with a reduction factor which reaches
$\eta\approx-0.057$. Within the plasma
model the reduction factor decays exponentially, while within the
two oscillator model it tends to a constant. This shows that the
repulsion at large distances does not have its origin in the
interaction between surface modes but stems from the repulsive
contribution of propagating modes.

The dotted lines are the short distance plasmon approximations to the model system
and for the realistic dielectric response functions. For the latter the plasmon approximation turns
out to be valid for distances up to twice the effective plasma
wavelength, that is about 170 nm. Indeed, the present calculation does not pretend
to be a precise description of the recent experiment~\cite{Capasso:2009}, but the
system $Au-C_6H_5Br-SiO_2$  is used as an illustration.

The calculation of the Casimir force at zero temperature for the
solid-liquid-solid system using measured dielectric functions of all
involved materials for the wavelength range from millimeters down to
subnanometers was carried out in~\cite{Palasantzas:2009}, which was
submitted simultaneously to the present paper.

\section{Conclusion}
In the present paper we have studied the Casimir force in the
systems  which meet the condition~(\ref{condition}).  The magnitude
of the force in the present setup is small. For example, in the
system $SiO_2-C_6H_5Br-Au$, with the 150 nm layer of bromobenzene, it is only
about 5\% of the force between gold covered plates in vacuum.

From previous studies~\cite{HenGreffet:2004, IntraLam:2005,
Bordag:2005, Pirozhenko:2008tr} we remember that when two plates in
vacuum form a cavity, the total energy of the coupled surface modes
is negative, corresponding to attraction. We have shown that within
the plasma model the surface modes repel at short distances when the
materials satisfy the condition~(\ref{condition}). Moreover the
Casimir repulsion is then completely due to the repulsion of the
surface modes. At medium and long distances the interaction of the
surface modes becomes attractive, but the dominating repulsive
contribution of the propagative modes leads to a total repulsive
force.

\acknowledgements We acknowledge financial support from the European
Contract No. STRP 12142 NANOCASE.


\begin{thebibliography}{1}
\bibitem{Lifshitz:Eng:1956}
E.~M. Lifshitz.
\newblock {\em Soviet \ Phys.\ JETP}, 2:73, 1956.

\bibitem{Lambrecht:2000}
A. Lambrecht and S. Reynaud.
\newblock{\em Eur. Phys. Journ.} \textbf{D8}, 309 (2000).

\bibitem{Boyer:1974}
T.~H. Boyer.
\newblock {\em Phys. Rev.} \textbf{A 9}, 2078--2084 (1974).


\bibitem{Kenneth:2002ij}
O.~Kenneth, I.~Klich, A.~Mann, and M.~Revzen.
 \newblock{\em Phys. Rev. Lett.} \textbf{89}, 033001 (2002).


\bibitem{Henkel:2005}
C. Henkel and K. Joulain.
\newblock {\em Europhys. Lett.} \textbf{9(6)}, 929--935 (2005).

\bibitem{Dalvit:2008}
F.S.S.~Rosa, D.A.R.~Dalvit, P.W.~Milonni .
\newblock{Phys. Rev. Lett.}\textbf{100}, 183602 (2008).

\bibitem{Pirozhenko:2008tr}
I.~G. Pirozhenko and A.~Lambrecht.
\newblock {\em Journal Physics} \textbf{A 41(16)}, 164015 (2008).

\bibitem{Veselago:2006}  V.G. Veselago, L. Braginsky, V. Shklover, Ch. Hefner,
\newblock {\em J. Comput. Theor. Nanosci.} \textbf{3}, N 2, 1 (2006).

\bibitem{Binns:2008} Chris Binns, Private communication.

\bibitem{Milling:1996}
A.~Milling, P.~Mulvaney, I.~Larson.
\newblock{\em J. of Colloid and Interface Science}, \textbf{180}, 460-465 (1996).

\bibitem{Lee:2001} Seung-woo Lee and Wolfgang M. Sigmund.
\newblock{\em J. of Colloid and Interface Science}, \textbf{243}, 365 - 369 (2001).

\bibitem{Lee:2002} Seung-woo Lee and Wolfgang M. Sigmund.
\newblock{\em Colloids and Surfaces A: Physicochemical and Engineering Aspects}, \textbf{204}, 43 - 50 (2002).

\bibitem{Rodriguez:2008}
Alejandro~W. Rodriguez, J.~N. Munday, J.~D. Joannopoulos, Federico Capasso,
  Diego A.~R. Dalvit, and Steven~G. Johnson.
\newblock {\em Phys. Rev. Lett.} \textbf{101}, 190404 (2008).

\bibitem{Munday:2007}
 J.~N. Munday and Federico Capasso.
  \newblock {\em Physical Review} \textbf{A 75}, 060102(R) (2007).

\bibitem{Munday:2008}
 J.~N. Munday, Federico Capasso, V. A. Parsegian and Sergey M. Bezrukov.
\newblock {\em Phys. Rev.} \textbf{A 78}, 032109 (2008).

\bibitem{Capasso:2007}
Federico Capasso, Jeremy~N. Munday, Davide Iannuzzi, and H.~B. Chan.
\newblock {\em IEEE JOURNAL OF SELECTED TOPICS IN QUANTUM ELECTRONICS},
  \textbf{13(2)} 400--414 (2007).


\bibitem{Palasantzas:2009a}
P.J. van Zwol, G. Palasantzas, J. Th. M. De Hosson.
\newblock{Weak dispersive forces between glass-gold macroscopic surfaces in alcohols} arXiv:0904.0622 (2009).

\bibitem{Capasso:2009} J.~N.~Munday, F.~Capasso, A.~Parsegian, \newblock {\em Nature} \textbf{457}, 170 (2009).


\bibitem{Bergstrom:1997}
Lennart Bergstr\"{o}m.
\newblock {\em Advances in Colloid and Interface Science} \textbf{70}, 125--169, 1997.

\bibitem{pirozhenko:013811}
I.~Pirozhenko and A.~Lambrecht.
\newblock {\em Physical Review} \textbf{A 77} 013811 (2008).

\bibitem{HenGreffet:2004}
C.~Henkel, K.~Joulain, J.-Ph. Mulet, and J.-J. Greffet.
\newblock {\em Phys. Rev.}\textbf{A 69} 023808 (2004).

\bibitem{IntraLam:2005}
F.~Intravaia and A.~Lambrecht.
\newblock {\em Physical Review Letters} \textbf{94} 110404 (2005); idem \textbf{96} 218902 (2006).

\bibitem{Bordag:2005}
M.~Bordag.
\newblock {\em Journal Physics} \textbf{A 39}, 6173--6186 (2005).

\bibitem{Palasantzas:2009}
P.J. van Zwol, G. Palasantzas, J. Th. M. De Hosson.
\newblock{The influence of dielectric properties on van der Waals/Casimir forces in
solid-liquid systems} arXiv:0905.0889 (2009)

\end{thebibliography}
\end{document}